\documentclass[conference]{IEEEtran}
\IEEEoverridecommandlockouts
\usepackage{algorithmic}
\usepackage{amsmath,amssymb,amsfonts,amstext,amsthm}
\usepackage{mathtools}\mathtoolsset{centercolon}
\usepackage{cite,dsfont,enumerate}
\usepackage[mathscr]{euscript}
\usepackage[geometry]{ifsym}
\usepackage[utf8]{inputenc} 
\usepackage[T1]{fontenc}
\usepackage{graphicx,hyperref,ifthen,makeidx,multicol}
\usepackage{pict2e}
\usepackage{shadethm}
\usepackage{textcomp,verbatim}
\usepackage{xcolor,xfrac,xr,url}
\usepackage{soul}
\usepackage{academicons}
\usepackage[keeplastbox]{flushend} 
\usepackage{balance} 
\definecolor{orcidlogocol}{HTML}{A6CE39}

\def\BibTeX{{\rm B\kern-.05em{\sc i\kern-.025em b}\kern-.08em
    T\kern-.1667em\lower.7ex\hbox{E}\kern-.125emX}}
\hypersetup{
	pdftitle		={On the Existence of the Augustin Mean},
	pdfauthor		={Baris Nakiboglu and Hao-Chung Cheng},
	pdfsubject		={ITW 2021},
	pdfkeywords		={Augustin information, Augustin mean}, 
}
\hypersetup{hidelinks}
\theoremstyle{plain}
\newtheorem{lemma}{Lemma}

\newtheorem*{conjecture*}{Conjecture}

\newtheorem{remark}{Remark}
\newtheorem*{remark*}{Remark}
\theoremstyle{definition}
\newtheorem{definition}{Definition} 

\newtheorem{example}{Example}
\definecolor{mygray}{gray}{0.4}
\newcommand{\set} [1]			{{\mathscr{{#1}}}}
\newcommand{\alg}[1]			{{\mathcal{{#1}}}}
\newcommand{\rndv}[1]      {{\mathsf{{#1}}}}
\newcommand{\oper}[1]      {{\mathtt{{#1}}}}
\newcommand{\msr}[1]       {{\it    {{#1}}}}
\newcommand{\cnst}[1]      {{\mathit{{#1}}}}

\newcommand{\integers}[1]	{{\mathbb{Z}}_{^{{#1}}}}

\newcommand{\reals}[1]		{{\mathbb{R}}_{^{{#1}}}}

\DeclareMathOperator*{\essup}{ess\,sup}

\newcommand{\dif}[1]       {{\mathrm{d}{#1}}}  
\newcommand{\der}[2]        {\tfrac{\dif{#1}}{\dif{#2}}}

\newcommand{\DEF}[0]			{{\!\!~\triangleq\!~}}  
\newcommand{\mtimes}[0]			{{\circledast}}

\newcommand{\AC}[0]            {{\prec}}  
  
\newcommand{\NAC}[0]           {{\nprec}}  
\newcommand{\abs}[1]           {{\left\lvert{{#1}}\right\lvert}}
\newcommand{\abp}[1]           {{\left\lvert{{#1}}\right\lvert^{+}}}

\newcommand{\lon}[1]           {\left\lVert{{#1}}\right\lVert} 
\newcommand{\nrm}[2]           {{\left\lVert{{#2}}\right\lVert}_{{#1}}}
\newcommand{\IND}[1]           {{\mathds{1}_{\{#1\}}}}

\newcommand{\blx}[0]           {{\cnst{n}}}

\newcommand{\domtr}[1]         {{\set{Q}}_{{#1}}}

\newcommand{\EXS}[2]         {{\bf E}_{{#1}}\!\left[{#2}\right]}

\newcommand{\EX}[1]          {\EXS{\!}{{#1}}}                      

\newcommand{\hqx}[2]    {{\omega}_{{#1}\!}\left({#2}\right)}
\newcommand{\hhx}[2]    {{\mathfrak{D}}_{{#1}\!}\left({#2}\right)}
\newcommand{\Bset}[3]   {{\set{B}}_{{#1}}^{{#2}}({#3})}   
\newcommand{\Lp}[2]          {{{\cnst{L}}}^{{#1}}({#2})}

\newcommand{\fX}[0]          {{\cnst{f}}}

\newcommand{\gX}[0]          {{\cnst{g}}}   
   
\newcommand{\hX}[0]          {{\cnst{h}}}

\newcommand{\RD}[3]				{{\cnst{D}}_{{#1}}            \!\left(\left.            \! {#2}\right\Vert {#3}                  \right)}

\newcommand{\CRD}[4]			{{\cnst{D}}_{{#1}}            \!\left(\left.\!\left.    \! {#2}\right\Vert {#3} \right\vert{{#4}}\right)}

\newcommand{\RMI}[3]			{{\cnst{I}}_{{#1}}            \!\left(                  \! {#2};         \!{#3}                \!\right)}

\newcommand{\Aop}[3]	{{\oper{T}}_{{#1},{#2}}\left({#3}\right)} 
 
\newcommand{\tAop}[4]	{{\oper{T}}_{{#1},{#2}}^{#3}\left({#4}\right)}

\newcommand{\rfm}[0]			{{{\msr{\nu}}}}

\newcommand{\rnb}[0]          {{\cnst{\beta}}}

\newcommand{\rno}[0]          {{\cnst{\alpha}}}

\newcommand{\oev}[0]           {{\set{E}}}

\newcommand{\fmea}[1]          {{{\alg{M}}^{^{+}}\!({#1})}}

\newcommand{\pmea}[1]          {{{\alg{P}}({#1})}}

\newcommand{\dinp}[0]          {{\cnst{x}}}
\newcommand{\inp}[0]           {{\rndv{X}}}
\newcommand{\inpS}[0]          {{\set{X}}}
\newcommand{\inpA}[0]          {{\alg{X}}}

\newcommand{\dout}[0]          {{\cnst{y}}}

\newcommand{\outS}[0]          {{\set{Y}}}
\newcommand{\outA}[0]          {{\alg{Y}}}

\newcommand{\sta}[0]           {{\rndv{Z}}}

\newcommand{\mA}[0]				{{\msr{a}}}

\newcommand{\mB}[0]				{{\msr{b}}}

\newcommand{\mP}[0]				{{\msr{p}}}

\newcommand{\mQ}[0]				{{\msr{q}}}    
\newcommand{\qmn}[1]			{{{\mQ}_{{#1}}}}

\newcommand{\mS}[0]				{{\msr{s}}}    
\newcommand{\smn}[1]			{{{\mS}_{{#1}}}}

\newcommand{\mW}[0]				{{\msr{w}}}

\newcommand{\Wm}[0]				{{{\cnst{W}}}}
\newcommand{\Wmn}[1]			{{{\cnst{W}}_{{#1}}}}
\newcommand{\Wma}[2]			{{{\cnst{W}}_{{#1}}^{{#2}}}}

\newcommand{\renyi}[0]								{R\'{e}nyi~}

\makeatletter
\DeclareRobustCommand{\bigplus}{%
	\mathop{\vphantom{\sum}\mathpalette\@bigplus\relax}\slimits@
}
\newcommand{\@bigplus}[2]{\vcenter{\hbox{\make@bigplus{#1}}}}
\newcommand{\make@bigplus}[1]{%
	\sbox\z@{$\m@th#1\sum$}%
	\setlength{\unitlength}{\wd\z@}%
	\begin{picture}(1.4,1.4)
	\linethickness{.17ex}
	\Line(.7,.14)(.7,1.26)
	\Line(.14,.7)(1.26,.7)
	\end{picture}%
}
\DeclareRobustCommand{\bigtimes}{%
	\mathop{\vphantom{\sum}\mathpalette\@bigtimes\relax}\slimits@
}
\newcommand{\@bigtimes}[2]{\vcenter{\hbox{\make@bigtimes{#1}}}}
\newcommand{\make@bigtimes}[1]{%
	\sbox\z@{$\m@th#1\sum$}%
	\setlength{\unitlength}{\wd\z@}%
	\begin{picture}(1,1)
	\linethickness{.17ex}
	\Line(.1,.1)(.9,.9)
	\Line(.1,.9)(.9,.1)
	\end{picture}%
}
\makeatother
\begin{document}
\title{\vspace{-0.1em}On the Existence of the Augustin Mean\vspace{-0.1em}
  \thanks{
	H.-C.~Cheng is supported by the Young Scholar Fellowship (Einstein Program) of the Ministry of Science and Technology in Taiwan (R.O.C.) under grant number MOST 109-2636-E-002-001 \& 110-2636-E-002-009, and is supported by the Yushan Young Scholar Program of the Ministry of Education in Taiwan (R.O.C.) under grant number NTU-109V0904 \& NTU-110V0904.}
\thanks{
	B.~Nakibo\u{g}lu is supported by  the Science Academy, Turkey, under The Science Academy's Young Scientist Award Program (BAGEP), and by the Scientific and Technological Research Council of Turkey (T\"{U}B\.{I}TAK) under Grant 119E053.}
}
\author{%
 \IEEEauthorblockN{Hao-Chung Cheng}
\IEEEauthorblockA{\textit{Department of Electrical Engineering and
}\\
\textit{Graduate Institute of Communication Engineering},\\
	\textit{Department of Mathematics, National Taiwan University}\\
	Taipei 10617, Taiwan (R.O.C.)\\
	\textit{Hon Hai (Foxconn) Quantum Computing Centre} \\
\href{https://orcid.org/0000-0003-4499-4679}{\includegraphics[scale=.3]{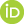} 0000-0003-4499-4679}\vspace{-2.6em}
}
\and
\IEEEauthorblockN{Bar\i\c{s} Nakibo\u{g}lu}
\IEEEauthorblockA{\textit{Department of Electrical and Electronics Engineering}\\
	\textit{Middle East Technical University}\\ 
	06800 Ankara, Turkey\\
\href{https://orcid.org/0000-0001-7737-5423}{\includegraphics[scale=.3]{Figures/orcid_24x24.png} 0000-0001-7737-5423}\vspace{-2em}
}
}
\maketitle

\begin{abstract}
The existence of a unique Augustin mean and its invariance under the Augustin operator 
are established for arbitrary input distributions with finite Augustin information 
for channels with countably generated output \(\sigma\)-algebras. 
The existence is established by representing the conditional Rényi divergence as a 
lower semicontinuous and convex functional in an appropriately chosen uniformly convex
space and then invoking the Banach--Saks property 
in conjunction with the lower semicontinuity and the convexity. 
A new family of operators is proposed to establish the invariance of the Augustin mean 
under the Augustin operator for orders greater than one. 
Some members of this new family strictly decrease the conditional Rényi divergence, 
when applied to the second argument of the divergence, 
unless the second argument is a fixed point of the Augustin operator.
\end{abstract}
\section{Introduction}
In sixties and seventies, Shannon's fundamental result has been strengthened
for memoryless channels in terms of three exponent functions:
\begin{enumerate}[(i)]
\item For codes operating at rates below the Shannon capacity,
the exponential decay rate of the error probability
with the block length is bounded from below by the random coding exponent \cite{gallager65,ebert66,richters67,augustin69,augustin78}
and from above by the sphere packing exponent \cite{shannonGB67A,haroutunian68,augustin69,augustin78}.
\item For codes operating at rates above the Shannon capacity, 
the exponential rate that the correct transmission (decoding) probability 
vanishes with  the block length is 
equal to the strong converse exponent, \cite{arimoto73,omura75,dueckK79}.
\end{enumerate}
These exponent functions have been characterized in terms of 
Gallager's functions \cite{gallager}, 
auxiliary channels \cite{csiszarkorner,csiszar98},
and Augustin information measures  \cite{augustin78}.
To obtain the right exponent functions for cost constrained codes
in terms of Gallager's functions, one has to apply the Lagrange multipliers method 
in a somewhat non-standard way described in \cite{gallager65,ebert66,richters67}.
The corresponding modification works for convex composition constraints, as well; see \cite{augustin78,scarlettMF14}.
This non-standard application of the Lagrange multipliers method to Gallager's function 
has recently been shown to be equivalent to the standard application 
of the Lagrange multipliers  method to the Augustin information measures in \cite[\S5]{nakiboglu19C}.
However, the Lagrange multipliers method is unnecessary to express the exponent functions 
in terms of Augustin information measures,  
either for composition constrained codes or for cost constrained codes. 
The right exponent functions are obtained by imposing the same constraints to the domain of 
the supremum defining Augustin capacity in terms of Augustin information
\cite{augustin78,poltyrev82,csiszar95,dalai17,dalaiW17,chengHT19,mosonyiO21,nakiboglu19C,nakiboglu20D,nakiboglu20F,chengN20A,verdu21}. 
Such characterizations permit relatively simple derivations of tight polynomial 
prefactors under certain symmetry hypothesis \cite{nakiboglu20F,chengN20A}.


Both the Augustin information and the \renyi information 
(i.e. a scaled and reparametrized version of Gallager's function \cite{nakiboglu19A}),
can be seen as generalizations of the mutual information. 
However, unlike the mutual information and the \renyi information,
the Augustin information does not have a closed form expression.
The order \(\rno\) Augustin information for the input distribution 
\(\mP\) is defined as\vspace{-.11cm}
\begin{align}
\label{eq:def:augustininformation}
\RMI{\rno}{\mP}{\Wm}
&\DEF \inf\nolimits_{\mQ\in\pmea{\outA}} \CRD{\rno}{\Wm}{\mQ}{\mP},
\end{align}~\vspace{-.52cm}\\ 
where \(\pmea{\outA}\) is the set of all probability measures on the output space.
For the case when the output set is a finite set 
(e.g.~when \(\Wm\) is a discrete memoryless channel as in \cite{csiszar95, verdu15}),
the compactness of \(\pmea{\outA}\),
the lower semicontinuity of \renyi divergence in its second argument \cite[Thm 15]{ervenH14},
and the extreme value theorem imply
the existence of an order \(\rno\) Augustin mean \(\qmn{\rno,\mP}\!\in\!\pmea{\outA}\) 
satisfying\vspace{-.11cm}
\begin{align}
\label{eq:def:augustinmean}
\RMI{\rno}{\mP}{\Wm}
&=\CRD{\rno}{\Wm}{\qmn{\rno,\mP}}{\mP}.
\end{align}~\vspace{-.52cm}\\ 
The Augustin mean \(\qmn{\rno,\mP}\) is unique because 
of  the strict convexity of 
the \renyi divergence in its second argument
described in \cite[Thm 12]{ervenH14}.
Other properties of the Augustin mean and information 
established in \cite{augustin78,nakiboglu19C}
can be derived independently, once the existence of a unique
Augustin mean  is established.

For channels whose output space is an arbitrary measurable space \((\outS,\outA)\), 
we no longer have the compactness of \(\pmea{\outA}\) and establishing the existence of the
Augustin mean becomes a more delicate issue. 
It has been established 
for the case when \(\mP\) is a probability mass function with a finite support set
for arbitrary channels in \cite{augustin78,nakiboglu19C}.
In addition, the closed form expression for the Augustin mean has been derived
for certain special cases:
for Gaussian input distributions on scalar or vector Gaussian channels in \cite{nakiboglu19C}
and for Augustin capacity achieving input distribution 
on additive exponential noise channels with a mean constraint in \cite{verdu21}. 
But a general existence result for the Augustin mean
has not been proved yet; see Remark~\ref{rem:definingAugustinMeanAsTheFixedPoint} 
of \S\ref{sec:fixedpoint} for a discussion regarding \cite{verdu21}.

In this paper, we prove, under finite Augustin information hypothesis, 
the existence of a unique Augustin mean,
its invariance under the Augustin operator,
and
its equivalence to the \(\qmn{\widetilde{\mP}}\) defined in \eqref{eq:def:outputdistributionAC},
which is absolute continuous in the output distribution 
\(\qmn{\mP}\) generated by the input distribution \(\mP\).
Our presentation will be as follows:
In \S\ref{sec:Preliminaries}, we introduce our model and notation
and prove that the infimum defining the Augustin information in \eqref{eq:def:augustininformation}
can be taken over the probability measures that are absolutely continuous 
in \(\qmn{\mP}\), rather than the whole \(\pmea{\outA}\).
In \S\ref{sec:existence}, we first use Radon--Nikodym theorem to express 
this optimization in \(\Lp{\tau}{\qmn{\mP}}\) for some \(\tau\!>\!1\),
with the help of a functional corresponding to the conditional \renyi divergence.
Then we show that this functional inherits the convexity and  the norm lower semicontinuity 
from the conditional \renyi divergence
and use them together with the Banach--Saks property
to establish the existence of a unique Augustin mean. 
In \S\ref{sec:fixedpoint}, we propose a new family of operators related to the Augustin operator,
establish a new monotonicity property 
for the conditional \renyi divergence, see Lemma \ref{lem:monotonicity},
and use it to establish the invariance of the Augustin mean under 
the Augustin operator.
In \S\ref{sec:discussions}, we briefly discuss the novelty of our approach
in comparison to the previous analysis methods, as we see it.

\section{Preliminaries}\label{sec:Preliminaries}
For any measurable space \((\outS,\outA)\), we denote the 
set of all probability measures on \((\outS,\outA)\) by
\(\pmea{\outA}\).
With a slight abuse of notation we denote 
the set of all probability measures 
that are absolutely continuous
with respect to a finite measure \(\mQ\) by 
\(\pmea{\mQ}\).
For finite measures, we use \(\fmea{\cdot}\) instead of \(\pmea{\cdot}\).
We use \(\lon{\cdot}\) for the total variation norm and corresponding metric.

\begin{definition}\label{def:divergence}
For any \(\rno\!\in\!(0,\infty] \), \(\mW\!\in\!\pmea{\outA}\),
 and \(\mQ\!\in\!\fmea{\outA}\)
\emph{the order \(\rno\) \renyi divergence between \(\mW\) and \(\mQ\)} 
is\vspace{-.1cm} 
\begin{align}
\notag
\!\!\RD{\rno}{\mW}{\mQ}\!
&\!\DEF\!
\begin{cases}
\tfrac{1}{\rno-1}\!\ln\!\int (\der{\mW}{\rfm})^{\rno} (\der{\mQ}{\rfm})^{1-\rno} \rfm(\dif{\dout})
&\rno\!\in\!\reals{+}\!\!\setminus\!\{1\}
\\
\int  \der{\mW}{\rfm}\left[ \ln\der{\mW}{\rfm} -\ln \der{\mQ}{\rfm}\right] \rfm(\dif{\dout})
&\rno=1
\\
\ln \essup_{\rfm} \der{\mW}{\rfm}/\der{\mQ}{\rfm}
&\rno\!=\!\infty
\end{cases}
\end{align}
where \(\rfm\) is any measure satisfying \(\mW\AC\rfm\) and \(\mQ\AC\rfm\).
\end{definition}
If \(\mQ\in\pmea{\outA}\), then 
\(\RD{\rno}{\mW}{\mQ}\) is positive unless \(\mW\!=\!\mQ\) by \cite[Thm. 8]{ervenH14}
and the following Pinsker's inequality holds by  \cite[Thms. 3 and 31]{ervenH14},\vspace{-.1cm}
\begin{align}
\label{eq:pinkser}
\RD{\rno}{\mW}{\mQ}
&\geq \tfrac{1\wedge \rno}{2} \lon{\mW-\mQ}^2
&
&\forall \mQ,\mW\in\pmea{\outA}.
\end{align}

We denote the set of all transition  probabilities\footnote{See \cite[Definition 9]{nakiboglu19A}, \cite[Definition 10.7.1]{bogachev} for the formal definition.}
from  \((\inpS,\inpA)\) to \((\outS,\outA)\) by \(\pmea{\outA|\inpA}\)
and model the channel \(\Wm\) as a transition  probability in \(\pmea{\outA|\inpA}\). 
Thus \cite[Thm. 10.7.2]{bogachev} ensures the existence of a joint distribution 
\(\mP\mtimes\Wm\)  on \(\inpA\otimes\outA\) 
for any input distribution \(\mP\) in \(\pmea{\inpA}\).
We call the \(\outA\)-marginal of \(\mP\mtimes\Wm\) 
the output distribution induced by \(\mP\) and denote it 
by \(\qmn{\mP}\).
\begin{align}
\label{eq:def:outputdistribution}
\qmn{\mP}(\oev)
&\DEF \mP\mtimes\Wm(\inpS\times\oev)
&
&\forall \oev\in\outA.
\end{align}
Applying \cite[Thm. 10.7.2]{bogachev} for \(\fX(\dinp,\dout)=\IND{\dout\in\oev}\) we get\vspace{-.1cm}
\begin{align}
\label{eq:outputdistribution}
\qmn{\mP}(\oev)
&=\int_{\inpS} \Wm(\oev|\dinp) \mP(\dif{\dinp}) 
&
&\forall \oev\in\outA.
\end{align}
With a slight abuse of notation,
for a \(\Wm\in\pmea{\outA|\inpA}\) and \(x\in\inpS\),
we denote the probability measure \(\Wm(\cdot|\dinp) \in \pmea{\outA}\) by 
\(\Wm(\dinp)\), whenever it is possible to do so without any ambiguity.

\begin{definition}\label{def:general-conditionaldivergence} 
	For any \(\rno\in(0,\infty]\), countably generated \(\sigma\)-algebra \(\outA\) 
	of subsets of \(\outS\), \(\Wm\!\in\!\pmea{\outA|\inpA}\), \(\mQ\!\in\!\fmea{\outA}\),
	and \(\mP\!\in\!\pmea{\inpA}\)
	\emph{the order \(\rno\) conditional \renyi divergence for the input distribution \(\mP\)} is
	\begin{align}
	\label{eq:def:general-conditionaldivergence}
	\CRD{\rno}{\Wm}{\mQ}{\mP}
	&\DEF \int \RD{\rno}{\Wm(\dinp)}{\mQ}\mP(\dif{\dinp}).
	\end{align}
\end{definition}
We assume \(\outA\) to be countably generated, so as to ensure 
the \(\inpA\)-measurablity of the integrand in \eqref{eq:def:general-conditionaldivergence} 
by\footnote{\cite[Lemma 37]{nakiboglu19C}
	establishes \(\inpA\)-measurability 
	for \(\mQ\!\in\!\pmea{\outA}\) and \(\rno\!\in\!\reals{+}\) case, but 
	a similar proof works for \(\mQ\!\in\!\fmea{\outA}\) and \(\rno\!\in\!(0,\infty]\) case.
} \cite[Lemma 37]{nakiboglu19C}.

For \(\rno=1\) case, one can confirm by substitution that 
the conditional \renyi divergence can be expressed in terms of 
the joint distribution \(\mP\mtimes\Wm\) induced by 
\(\mP\in\pmea{\inpA}\) as follows
\begin{align}
\label{eq:presibson-one}
\hspace{-.15cm}
\CRD{1}{\Wm}{\mQ}{\mP}
&\!=\!\RD{1}{\mP\mtimes\Wm}{\mP\otimes\mQ}~~~
&
&~~~~~\forall\mQ\!\in\!\pmea{\outA},
\end{align}
where \(\mP\otimes\mQ\) is the product measure.
Furthermore, \eqref{eq:outputdistribution} and  \eqref{eq:presibson-one}
can be used to confirm by substitution that
\begin{align}
\label{eq:sibson-one}
\hspace{-.15cm}
\CRD{1}{\Wm}{\mQ}{\mP}
&\!=\!\CRD{1}{\Wm}{\qmn{\mP}}{\mP}\!+\!\RD{1}{\qmn{\mP}}{\mQ}
&
&~\forall\mQ\!\in\!\pmea{\outA}.
\end{align}

\begin{definition}\label{def:general-information}
	For any \(\rno\in(0,\infty]\), countably generated \(\sigma\)-algebra \(\outA\), 
	\(\Wm\in\pmea{\outA|\inpA}\), and  \(\mP\in\pmea{\inpA}\)
	\emph{the order \(\rno\) Augustin information for the input distribution \(\mP\)}
	is	given by \eqref{eq:def:augustininformation}.
\end{definition}

For \(\rno\!=\!1\) case, \eqref{eq:sibson-one} provides us a closed form expression 
of the Augustin information by \eqref{eq:pinkser}: 
\(\RMI{1}{\mP}{\Wm}\!=\!\CRD{1}{\Wm}{\qmn{\mP}}{\mP}\).
For other orders, however,  a general closed form expression does not exist 
either for the Augustin information or for  the probability measure
that achieves the infimum given in \eqref{eq:def:augustininformation},
called the Augustin mean. 
Nevertheless \(\qmn{\mP}\),
can be used to restrict the domain of the optimization  
problem defining Augustin information as follows.

\begin{lemma}\label{lem:AIDequivalence}
For any \(\rno\in(0,\infty]\), countably generated \(\sigma\)-algebra \(\outA\), 
\(\Wm\in\pmea{\outA|\inpA}\), and  \(\mP\in\pmea{\inpA}\),
\begin{align}
\label{eq:lem:AIDequivalence}
\RMI{\rno}{\mP}{\Wm}
&=\inf\nolimits_{\mQ\in\pmea{\qmn{\mP}}} \CRD{\rno}{\Wm}{\mQ}{\mP}.
\end{align} 
\end{lemma}
\begin{proof}
Any \(\mQ\in\pmea{\outA}\) can be written as the sum
of absolutely continuous and singular components with respect to
\(\qmn{\mP}\)  by the Lebesgue decomposition theorem
\cite[Thm. 3.2.3]{bogachev}, i.e. there exist
\(\qmn{\sim}\!\AC\!\qmn{\mP}\) and 
\(\qmn{\perp}\!\perp\!\qmn{\mP}\)
such that \(\mQ\!=\!\qmn{\sim}\!+\!\qmn{\perp}.\)
Hence, there exists an \(\oev\!\in\!\outA\) satisfying 
\(\qmn{\mP}(\oev)\!=\!0\) and \(\qmn{\perp}(\outS\setminus\oev)\!=\!0\)
because \(\qmn{\perp}\!\perp\!\qmn{\mP}\).
Then \(\Wm(\oev|\dinp)=0\) \(\mP\)-a.s.~by \eqref{eq:outputdistribution}
and consequently
\begin{align}
\notag
\RD{\rno}{\Wm(\dinp)}{\mQ}
&=\RD{\rno}{\Wm(\dinp)}{\qmn{\sim}}
&
&\mP\text{-a.s.}
\end{align}
Thus \(\lon{\qmn{\sim}}\!>\!0\) for all \(\mQ\) satisfying \(\CRD{\rno}{\Wm}{\mQ}{\mP}\!<\!\infty\) and 
\begin{align}
\label{eq:absolutelycontinuouspart}
\CRD{\rno}{\Wm}{\mQ}{\mP}
&=\CRD{\rno}{\Wm}{\tfrac{\qmn{\sim}}{\lon{\qmn{\sim}}}}{\mP}
-\ln \lon{\qmn{\sim}}
\end{align}
for all \(\mQ\!\in\!\pmea{\outA}\) satisfying \(\CRD{\rno}{\Wm}{\mQ}{\mP}\!<\!\infty\).
Then we can replace \(\pmea{\outA}\) with \(\pmea{\qmn{\mP}}\) in \eqref{eq:def:augustininformation},
without changing the value of the infimum
because \(-\ln \lon{\qmn{\sim}}\!\geq\!0\)
and \(\tfrac{\qmn{\sim}}{\lon{\qmn{\sim}}}\!\in\!\pmea{\qmn{\mP}}\).
\end{proof}

\section{Existence of a Unique Augustin Mean}\label{sec:existence}

The uniform convexity\footnote{Usually, \(\mP\) rather than \(\tau\) is used to
	name the norm and the associated Banach space. 
	We deviate from the convention to reserve the symbol \(\mP\) for the input distributions.} 
of \(L^{\tau}\) for \(\tau>1\),
plays a central role in our proof of the
existence of a unique Augustin mean for input distributions
with finite Augustin information.
Let us first recall the definition of 
the \(\tau\)-norm.
For any \(\tau\geq 1\) and \(\qmn{\mP}\)-measurable function \(\fX:\outS\to\reals{}\),
the \(\tau\)-norm of \(\fX\) is\vspace{-.11cm}
\begin{align}
\label{eq:def:norm}
\nrm{\tau}{\fX}
&\DEF \left(\int \abs{\fX(\dout)}^{\tau} \qmn{\mP}(\dif{\dout})\right)^{\sfrac{1}{\tau}}.
\end{align}
The set of all finite \(\tau\)-norm functions \(\Lp{\tau}{\qmn{\mP}}\) 
form a complete normed vector space, i.e.~Banach space, under the pointwise addition 
and the scalar multiplication by \cite[Thm. 4.1.3]{bogachev}\vspace{-.11cm}
\begin{align}
\label{eq:def:Lp}
\Lp{\tau}{\qmn{\mP}}
&\DEF\left\{\fX: \nrm{\tau}{\fX}<\infty\right\}.
\end{align}
As a result of Radon--Nikdoym theorem \cite[Thm. 3.2.2]{bogachev},
we know that elements of \(\pmea{\qmn{\mP}}\) can be represented
via their Radon--Nikodym derivatives with respect to \(\qmn{\mP}\),
which will be non-negative functions of unit norm in \(\Lp{1}{\qmn{\mP}}\). 
By taking pointwise \(\tau^{\text{th}}\) root of these Radon--Nikodym derivatives, 
we can obtain analogous representations in \(\Lp{\tau}{\qmn{\mP}}\) 
for any positive \(\tau\). 
Motivated by these observations we define the following
subsets of \(\Lp{\tau}{\qmn{\mP}}\):\vspace{-.11cm}
\begin{align}
\label{eq:def:positivefunctions}
\Bset{}{\tau}{\qmn{\mP}}
&\DEF\{\fX\in \Lp{\tau}{\qmn{\mP}}:\fX(\dout)\geq 0~\qmn{\mP}\text{-a.s.}\},
\\
\label{eq:def:unitnormpositivefunctions}
\Bset{1}{\tau}{\qmn{\mP}}
&\DEF\{\fX\in \Bset{}{\tau}{\qmn{\mP}}:\nrm{\tau}{\fX}=1 \},
\\
\label{eq:def:subunitnormpositivefunctions}
\Bset{\leq 1}{\tau}{\qmn{\mP}}
&\DEF\{\fX\in \Bset{}{\tau}{\qmn{\mP}}:\nrm{\tau}{\fX}\leq 1 \}.
\end{align}
Let \(\hqx{\tau}{\cdot}:\Bset{}{\tau}{\qmn{\mP}}\to \fmea{\qmn{\mP}}\) 
be the function defined through the following relation
\vspace{-.11cm}
\begin{align}
\label{eq:def:inverseRND}
\!\hqx{\tau}{\fX}(\oev)
&\!\DEF\!\!\!\int_{\oev}[\fX(\dout)]^\tau \qmn{\mP}(\dif{\dout})
&&&
&\forall\fX\!\in\!\Bset{}{\tau}{\qmn{\mP}},\oev\!\in\!\outA. 
\end{align}
Using the conditional \renyi divergence and \(\hqx{1}{\cdot}\), 
we can define the functional 
\(\CRD{\rno}{\Wm}{\hqx{1}{\cdot}}{\mP}\) on \(\Bset{}{1}{\qmn{\mP}}\),
which inherits the convexity and  norm lower semicontinuity 
from the \renyi divergence by the linearity and continuity 
of \(\hqx{1}{\cdot}\).
Lemmas \ref{lem:inherited_convexity} and \ref{lem:inherited_lsc}
demonstrate that for an appropriately chosen \(\tau\!>\!1\),
the functional 
\(\CRD{\rno}{\Wm}{\hqx{\tau}{\cdot}}{\mP}\) on \(\Bset{}{\tau}{\qmn{\mP}}\)
inherits the convexity and  norm lower semicontinuity, as well.
These observations are important because, unlike \(\Lp{1}{\qmn{\mP}}\), 
\(\Lp{\tau}{\qmn{\mP}}\) is uniformly convex for any \(\tau\!>\!1\), 
and thus it has the Banach--Saks property.\vspace{-.04cm}
\begin{definition}
Let \(\hhx{\rno}{\cdot}:\Bset{}{\tau_{\rno}}{\qmn{\mP}}\!\to\!(-\infty,\infty]\) 
be\vspace{-.11cm}
\begin{align}
\label{eq:def:functional}
\hhx{\rno}{\fX}
&\!\DEF\!\CRD{\rno}{\Wm}{\hqx{\tau_{\rno}}{\fX}}{\mP} 
&
&
\end{align}
for all \(\fX\in \Bset{}{\tau_{\rno}}{\qmn{\mP}}\) and \(\rno\!\in\!(0,\infty]\), 
where\vspace{-.11cm}
\begin{align}
\label{eq:def:orderofthefunctional}
\tau_{\rno}
&\DEF
\begin{cases}
2
&\rno\in[0.5,\infty]
\\
\frac{1}{1-\rno}
&\rno\in(0,0.5)
\end{cases}.
\end{align}
\end{definition}
\begin{lemma}\label{lem:inherited_convexity}
	For	all \(\rno\!\in\!(0,\infty]\), functional 
	\(\hhx{\rno}{\cdot}\), defined in \eqref{eq:def:functional},
	is convex on \(\Bset{}{\tau_{\rno}}{\qmn{\mP}}\).	
\end{lemma}
\begin{lemma}\label{lem:inherited_lsc}
	For all \(\rno\!\in\!(0,\infty]\), functional 
	\(\hhx{\rno}{\cdot}\), defined in \eqref{eq:def:functional},
	is norm lower semicontinuous on \(\Bset{}{\tau_{\rno}}{\qmn{\mP}}\).	
\end{lemma}
\noindent Proofs of Lemmas \ref{lem:inherited_convexity} and \ref{lem:inherited_lsc}
are presented in Appendix \ref{sec:convexity} and Appendix \ref{sec:lsc}.

\begin{lemma}\label{lem:existence}
For all \(\rno\!\in\!(0,\infty]\), 
there exists an \(\fX_{\rno}\in\Bset{}{\tau_{\rno}}{\qmn{\mP}}\)
satisfying \(\nrm{\tau_{\rno}}{\fX_{\rno}}=1\) and
\vspace{-.1cm}
\begin{align}
\label{eq:lem:existence}
\hhx{\rno}{\fX_{\rno}}
&=\RMI{\rno}{\mP}{\Wm}.
\end{align}
\end{lemma}
\begin{proof}
Note that \(\hqx{\tau}{\gamma\fX}=\gamma^{\tau}\hqx{\tau}{\fX}\) 
for all \(\tau\geq 1\)
and \(\gamma\geq 0\)
by \eqref{eq:def:inverseRND}.
Thus
\vspace{-.1cm}
\begin{align}
\label{eq:existence1}
\hhx{\rno}{\sfrac{\fX}{\nrm{\tau_{\rno}}{\fX}}}
&=\hhx{\rno}{\fX}+\ln\nrm{\tau_{\rno}}{\fX}^{\tau_{\rno}}.
\end{align}
for all \(\fX\in\Bset{}{\tau_{\rno}}{\qmn{\mP}}\) by \eqref{eq:def:functional}.
Consequently, 
\vspace{-.1cm}
\begin{align}
\notag
\inf\nolimits_{\fX\in\Bset{\leq 1}{\tau_{\rno}}{\qmn{\mP}}}
\hhx{\rno}{\fX}
&=\inf\nolimits_{\fX\in\Bset{1}{\tau_{\rno}}{\qmn{\mP}}}
\hhx{\rno}{\fX}.
\end{align}
Hence the definition of \(\hhx{\rno}{\cdot}\), the Radon--Nikdoym theorem \cite[Thm 3.2.2]{bogachev},
and Lemma \ref{lem:AIDequivalence} imply
\vspace{-.1cm}
\begin{align}
\label{eq:existence2}
\inf\nolimits_{\fX\in\Bset{\leq 1}{\tau_{\rno}}{\qmn{\mP}}}
\hhx{\rno}{\fX}
&=\RMI{\rno}{\mP}{\Wm}.
\end{align}
Thus there exists a sequence \(\{\fX_{\blx}\}\subset\Bset{\leq 1}{\tau_{\rno}}{\qmn{\mP}}\) 
satisfying\footnote{For example let \(\fX_{\blx}\) be such that
\(\hhx{\rno}{\fX_{\blx}}\!\leq\!\RMI{\rno}{\mP}{\Wm}\!+\!(\sfrac{1}{\blx})^2\).}
\vspace{-.1cm}
\begin{align}
\label{eq:existence3}
\hhx{\rno}{\fX_{\blx}}
&\downarrow \RMI{\rno}{\mP}{\Wm},
\\
\label{eq:existence4}
\sum\nolimits_{\blx\in\integers{+}}[\hhx{\rno}{\fX_{\blx}}-\RMI{\rno}{\mP}{\Wm}] 
&<\infty.
\end{align}
\(\Lp{\tau_{\rno}}{\qmn{\mP}}\) has the Banach--Saks property for \(\tau_{\rno}\in (1,2]\) by \cite[Cor.~4.7.17]{bogachev},
because it is uniformly convex by \cite[Thm. 4.7.15]{bogachev}. 
Thus
for the norm bounded sequence \(\{\fX_{\blx}\}\),
there exist a subsequence \(\{\fX_{\blx_{k}}\}\)
and an \(\fX_{\rno} \in \Lp{\tau_{\rno}}{\qmn{\mP}}\) such that 
\vspace{-.1cm}
\begin{align}
\label{eq:existence5}
\lim\nolimits_{k\to\infty} \nrm{\tau_{\rno}}{\tfrac{\fX_{\blx_{1}}+\cdots+\fX_{\blx_{k}}}{k}-\fX_{\rno}}
&=0.
\end{align}
Furthermore, \(\fX_{\rno}\!\in\!\Bset{\leq 1}{\tau_{\rno}}{\qmn{\mP}}\) because 
\(\Bset{\leq 1}{\tau_{\rno}}{\qmn{\mP}}\) is closed and
\(\tfrac{\fX_{\blx_{1}}+\cdots+\fX_{\blx_{k}}}{k}\!\in\!\Bset{\leq 1}{\tau_{\rno}}{\qmn{\mP}}\)
for all \(k\) 
by  the non-negativity of \(\fX_{\blx}\)'s 
and the triangle inequality of \(\nrm{\tau_{\rno}}{\cdot}\).

\noindent The norm lower semicontinuity of \(\hhx{\rno}{\cdot}\)
established in Lemma \ref{lem:inherited_lsc},
\(\fX_{\rno}\!\in\!\Bset{\leq 1}{\tau_{\rno}}{\qmn{\mP}}\),
and \eqref{eq:existence5} imply
\vspace{-.1cm}
\begin{align}
\label{eq:existence6}
\hhx{\rno}{\fX_{\rno}}
&\leq \liminf\nolimits_{k\to\infty} \hhx{\rno}{\tfrac{\fX_{\blx_{1}}+\cdots+\fX_{\blx_{k}}}{k}}.
\end{align}
On the other hand, the convexity of \(\hhx{\rno}{\cdot}\)
established in Lemma \ref{lem:inherited_convexity} implies
\vspace{-.1cm}
\begin{align}
\label{eq:existence7}
\hhx{\rno}{\tfrac{\fX_{\blx_{1}}+\cdots+\fX_{\blx_{k}}}{k}}
&\leq 
\tfrac{\hhx{\rno}{\fX_{\blx_{1}}}+\cdots+\hhx{\rno}{\fX_{\blx_{k}}}}{k}.
\end{align}
\(\hhx{\rno}{\fX_{\rno}}\leq \RMI{\rno}{\mP}{\Wm}\)
by \eqref{eq:existence3}, \eqref{eq:existence4}, \eqref{eq:existence6} and \eqref{eq:existence7}.
Hence, \eqref{eq:lem:existence} follows from 
\eqref{eq:existence2}
and the fact that \(\fX_{\rno}\!\in\!\Bset{\leq 1}{\tau_{\rno}}{\qmn{\mP}}\).
Furthermore,
\(\nrm{\tau_{\rno}}{\fX_{\rno}}\!=1\) 
as a result of
\eqref{eq:lem:existence},
\eqref{eq:existence1},
and
\eqref{eq:existence2}.
\end{proof}
For finite orders, Lemma \ref{lem:Augustinmeanexistence},
expresses Lemma \ref{lem:existence} 
in terms of probability measures and strengthens it with 
uniqueness assertion 
for the finite Augustin information case. 

\begin{lemma}\label{lem:Augustinmeanexistence}
	For any \(\rno\!\in\!\reals{+}\), 
	channel \(\Wm\!\in\!\pmea{\outA|\inpA}\) with a countably generated output \(\sigma\)-algebra \(\outA\), 
	and  input distribution \(\mP\!\in\!\pmea{\inpA}\)
	with a finite \emph{order \(\rno\) Augustin information},
there exists a unique \(\qmn{\rno,\mP}\!\in\!\pmea{\outA}\) satisfying\vspace{-.1cm}
\begin{align}
\label{eq:lem:Augustinmeanexistence}
\RMI{\rno}{\mP}{\Wm}\!=\!\CRD{\rno}{\Wm}{\qmn{\rno,\mP}}{\mP},
\end{align}
called the order \(\rno\) Augustin mean for the input distribution \(\mP\).
Furthermore, \(\qmn{\rno,\mP}\) is absolutely continuous in \(\qmn{\mP}\), i.e. \(\qmn{\rno,\mP}\AC\qmn{\mP}\).
\end{lemma}
\noindent Proof of Lemma \ref{lem:Augustinmeanexistence}
is presented in the Appendix \ref{sec:ExitenceProof}.

\section{Fixed Point Properties of Augustin Mean}\label{sec:fixedpoint} 
The existence of a unique Augustin mean \(\qmn{\rno,\mP}\) 
and its absolute continuity in \(\qmn{\mP}\) 
are important observations. But they do not provide an easy way to 
decide whether \(\qmn{\rno,\mP}\!=\!\mQ\) for a \(\mQ\AC\qmn{\mP}\) or not. 
For input distributions that are probability mass functions 
with finite support set, this issue was addressed by characterizing 
\(\qmn{\rno,\mP}\)
as the only fixed point of the Augustin operator that is equivalent to \(\qmn{\mP}\), 
see\footnote{This is the case even for certain quantum models \cite[Proposition 4]{chengGH19}.}
\cite[Lemma 34.2]{augustin78}, \cite[Lemma 13]{nakiboglu19C}.
Our main goal in this section is to establish an analogous characterization 
of the Augustin mean \(\qmn{\rno,\mP}\) for a general input distribution \(\mP\) 
merely by assuming that \(\RMI{\rno}{\mP}{\Wm}\)
is finite, see Lemma \ref{lem:fixedpoint}. 
Let \(\domtr{\rno,\mP}\), \(\inpS_{\rno,\mP}^{\mQ}\), and \(\inpA_{\rno,\mP}^{\mQ}\) be
\vspace{-.1cm}
\begin{align}
\notag\domtr{\rno,\mP}
&\DEF\{\mQ\!\in\!\pmea{\outA}:\CRD{\rno}{\Wm}{\mQ}{\mP}\!<\!\infty\},
\\
\notag\inpS_{\rno,\mP}^{\mQ}
&\DEF\{\dinp:\RD{\rno}{\Wm(\dinp)}{\mQ}<\infty\},
\\
\notag
\inpA_{\rno,\mP}^{\mQ}
&\DEF\{\oev\cap \inpS_{\rno,\mP}^{\mQ}:\oev\in\inpA\}.
\end{align}
\begin{definition}\label{def:tiltedchannel}
For any \(\rno\in\reals{+}\), countably generated \(\sigma\)-algebra \(\outA\) 
of subsets of \(\outS\), \(\Wm\in\pmea{\outA|\inpA}\), \(\mQ\in\pmea{\outA}\),
and \(\dinp\in\inpS_{\rno,\mP}^{\mQ}\)
\vspace{-.1cm}
\begin{align}
\label{eq:def:tiltedchannel}
\der{\Wma{\rno}{\mQ}(\dinp)}{\rfm}
&\DEF e^{(1-\rno)\RD{\rno}{\Wm(\dinp)}{\mQ}}
\left(\der{\Wm(\dinp)}{\rfm}\right)^{\rno} 
\left(\der{\mQ}{\rfm}\right)^{1-\rno}.
\end{align}
Then \(\Wma{\rno}{\mQ}\) defines a transition probability called 
\emph{the order \(\rno\) tilted channel}  \(\Wma{\rno}{\mQ}\in\pmea{\outA|\inpA_{\rno,\mP}^{\mQ}}\).
\end{definition}
\begin{remark}
If \(\mQ\in\domtr{\rno,\mP}\), then \(\mP(\inpS_{\rno,\mP}^{\mQ})\!=\!1\).
Hence, for input distributions that are absolutely continuous in \(\mP\),
the fact that \(\Wma{\rno}{\mQ}\) is an element of 
\(\pmea{\outA|\inpA_{\rno,\mP}^{\mQ}}\) rather than \(\pmea{\outA|\inpA}\)
is inconsequential. 
\end{remark}
\begin{definition}\label{def:preliminary}
Under the hypothesis of Lemma \ref{lem:Augustinmeanexistence},	
\emph{the Augustin operator} 
\(\Aop{\rno}{\mP}{\cdot}\!:\!\domtr{\rno,\mP}\!\to\!\pmea{\outA}\) 
is defined as
\vspace{-.1cm}
	\begin{align}
	\label{eq:def:Aoperator}
	\Aop{\rno}{\mP}{\mQ}(\oev)
	&\!\DEF\!\EXS{\mP\!}{\Wma{\rno}{\mQ}(\oev|\inp)}
	&
	&\forall \oev\!\in\!\outA,~\mQ\!\in\!\domtr{\rno,\mP}.
	\end{align}
Furthermore, for any \(\rnb\in\reals{+}\) satisfying
	\(\RD{\rnb}{\Aop{\rno}{\mP}{\mQ}}{\mQ}<\infty\),
	\emph{the tilted Augustin operator}	\(\tAop{\rno}{\mP}{\rnb}{\mQ}\) is defined as
	\begin{align}
	\label{eq:def:tiltedAoperator}
	\der{\tAop{\rno}{\mP}{\rnb}{\mQ}}{\rfm}
	&\DEF 
	e^{(1-\rnb)\RD{\rnb}{\Aop{\rno}{\mP}{\mQ}}{\mQ}}
	\left(\der{\Aop{\rno}{\mP}{\mQ}}{\rfm}\right)^{\rnb}
	\left(\der{\mQ}{\rfm}\right)^{1-\rnb}.
	\end{align}
\end{definition}
The Augustin operator has been used before 
either implicitly \cite{fano,haroutunian68,poltyrev82}
or explicitly \cite{augustin78,nakiboglu19C,chengGH19,verdu21}.
However, to the best of our knowledge, the tilted
Augustin operator is first defined and analyzed in the present work.
\begin{lemma}\label{lem:monotonicity}
Under the hypothesis of Lemma \ref{lem:Augustinmeanexistence},
if either \(\rno\in(0,1)\) and \(\rnb\!\in\!(0,1]\),
or \(\rno\in(1,\infty)\) and  \(\rnb\!\in\!(0,1\wedge\frac{1}{\rno-1})\),
then for any \(\mQ\!\in\!\domtr{\rno,\mP}\) we have
\vspace{-.1cm}
\begin{align}
\notag
\hspace{1.2cm}&\hspace{-1.2cm}
\CRD{\rno}{\Wm}{\mQ}{\mP}
\!-\!
\CRD{\rno}{\Wm}{\tAop{\rno}{\mP}{\rnb}{\mQ}}{\mP}
\\
\notag
&\geq 
\rnb\RD{1-\rnb\abp{\rno-1}}{\Aop{\rno}{\mP\!}{\mQ}}{\!\mQ}
\!+\!(1\!-\!\rnb)\!\RD{\rnb}{\Aop{\rno}{\mP\!}{\mQ}}{\!\mQ}
\\
\notag
&\geq \tfrac{\rnb(2-\rnb(\rno\vee 1))}{2}
\lon{\Aop{\rno}{\mP}{\mQ}-\mQ}^2.
\end{align}
\end{lemma}
A particular case of Lemma \ref{lem:monotonicity} for \(\rno\!\in\!(0,1)\) and \(\rnb\!=\!1\) 
was proved in \cite[p. 236]{augustin78} and \cite[(B.4)]{nakiboglu19C}, 
and was used to show that the Augustin mean is a fixed point of the Augustin 
operator\footnote{Although we will not rely on it, it is worth mentioning that \(\tAop{\rno}{\mP}{\rnb}{\mQ}=\mQ\) 
	holds either for all positive real \(\rnb\)'s or for none.}
in \cite[Lemma 34.2]{augustin78} and \cite[Lemma 13 (c)]{nakiboglu19C}
for \(\rno\!\in\!(0,1)\).
Lemma \ref{lem:monotonicity} allows us to invoke this simpler argument for 
establishing the fixed point property for \(\rno\!\in\!(1,\infty)\) case.
\begin{proof}\vspace{-.1cm}
\begin{align}
\notag
\hspace{0.1cm}&\hspace{-0.2cm}
\CRD{\rno}{\Wm}{\mQ}{\mP}
\!-\!
\CRD{\rno}{\Wm}{\tAop{\rno}{\mP}{\rnb}{\mQ}}{\mP}
\\
\notag
&=\tfrac{1}{1-\rno}\EXS{\mP\!}{\ln \int  \left(\der{\tAop{\rno}{\mP}{\rnb}{\mQ}}{\mQ}\right)^{1-\rno}\Wma{\rno}{\mQ}(\dif{\dout}|\inp)}
\\
\notag
&\stackrel{(a)}{\geq} 
\begin{cases}
\tfrac{1}{1-\rno} \EXS{\mP\!}{\displaystyle{\int}  \ln \left(\der{\tAop{\rno}{\mP}{\rnb}{\mQ}}{\mQ}\right)^{1-\rno}\Wma{\rno}{\mQ}(\dif{\dout}|\inp)}
&\text{if~}\rno\!<\!1
\\
\tfrac{1}{1-\rno}\ln\EXS{\mP\!}{~\displaystyle{\int} \left(\der{\tAop{\rno}{\mP}{\rnb}{\mQ}}{\mQ}\right)^{1-\rno}\Wma{\rno}{\mQ}(\dif{\dout}|\inp)}
&\text{if~}\rno\!>\!1
\end{cases}
\\
\notag
&\stackrel{(b)}{=}
\begin{cases}
\displaystyle{\int}
\der{\Aop{\rno}{\mP}{\mQ}}{\mQ} \ln\left(\der{\tAop{\rno}{\mP}{\rnb}{\mQ}}{\mQ}\right)
\mQ(\dif{\dout})
&\text{if~}\rno\!<\!1
\\
\tfrac{1}{1-\rno} \ln \displaystyle{\int} \left(\der{\tAop{\rno}{\mP}{\rnb}{\mQ}}{\mQ}\right)^{1-\rno}
\der{\Aop{\rno}{\mP}{\mQ}}{\mQ}
\mQ(\dif{\dout})
&\text{if~}\rno\!>\!1
\end{cases}
\\
\notag
&\stackrel{(c)}{=}
\begin{cases}
\rnb\RD{1}{\Aop{\rno}{\mP\!}{\mQ}}{\!\mQ}
\!+\!(1\!-\!\rnb)\!\RD{\rnb}{\Aop{\rno}{\mP\!}{\mQ}}{\!\mQ}
&\text{if~}\rno\!<\!1
\\
\rnb\RD{1+\rnb(1-\rno)}{\Aop{\rno}{\mP\!}{\mQ}}{\!\mQ}
\!+\!(1\!-\!\rnb)\!\RD{\rnb}{\Aop{\rno}{\mP\!}{\mQ}}{\!\mQ}
&\text{if~}\rno\!>\!1
\end{cases}
\end{align}
where \((a)\) follows from Jensen's inequality and
the concavity of natural logarithm function,
\((b)\) follows from \eqref{eq:def:Aoperator}
and Fubini's theorem \cite[Thm. 3.4.4]{bogachev},
\((c)\) follows \eqref{eq:def:tiltedAoperator}.
The second inequality of the lemma follows from \eqref{eq:pinkser}. 
\end{proof}
For most, but not all, cases of interest \(\Wm(\dinp)\AC\qmn{\mP}\) \(\mP\)-a.s.,
e.g.~see Example \ref{eg:partiallynoiseless}.
To avoid introducing ``\(\Wm(\dinp)\AC\qmn{\mP}\) \(\mP\)-a.s.'' 
as a separate hypothesis, we define \(\qmn{\widetilde{\mP}}\) as follows
\begin{align}
\label{eq:def:outputdistributionAC}
\der{\qmn{\widetilde{\mP}}}{\qmn{\mP}}
&\DEF\EXS{\mP\!}{\der{\Wmn{\sim}(\inp)}{\qmn{\mP}}},
\end{align}
where \(\Wmn{\sim}(\dinp)\) is the \(\qmn{\mP}\)-absolutely continuous part of \(\Wm(\dinp)\).
Note that \(\Wmn{\sim}(\dinp)\!=\!\Wm(\dinp)\) \(\mP\)-a.s.~and  thus \(\qmn{\widetilde{\mP}}\!=\!\qmn{\mP}\)
whenever \(\Wm(\dinp)\AC\qmn{\mP}\) \(\mP\)-a.s.~and thus whenever \(\RMI{1}{\mP}{\Wm}<\infty\).

\begin{lemma}\label{lem:fixedpoint}
	Under the hypothesis of Lemma \ref{lem:Augustinmeanexistence},
	\(\qmn{\rno,\mP}\sim \qmn{\widetilde{\mP}}\),\vspace{-.1cm}
	\begin{align}
	\label{eq:lem:fixedpoint}
	\Aop{\rno}{\mP}{\qmn{\rno,\mP}}
	&=\qmn{\rno,\mP},
	\\
	\label{eq:lem:EHB}
	\hspace{-.3cm}\RD{1\vee \rno}{\qmn{\rno,\mP}}{\!\mQ}\!\geq\! 
	\CRD{\rno}{\Wm}{\!\mQ}{\!\mP}\!-\!\RMI{\rno}{\mP}{\Wm}
	&\!\geq\!\RD{1 \wedge \rno}{\qmn{\rno,\mP}}{\!\mQ}\!,
	\end{align}
	for all \(\mQ\!\in\!\pmea{\outA}\).
	Furthermore, if \(\qmn{\widetilde{\mP}}\AC \mQ\) and \(\Aop{\rno}{\mP}{\mQ}\!=\!\mQ\)
	for a \(\mQ\in\pmea{\outA}\), then \(\qmn{\rno,\mP}\!=\!\mQ\).
\end{lemma}

\begin{remark}
	For \(\rno\!\in\!(1,\infty)\), 
	\(\CRD{\rno}{\Wm}{\mQ}{\mP}\) is finite and \(\Aop{\rno}{\mP}{\mQ}\) is defined
	only for \(\mQ\)'s satisfying \(\qmn{\mP}\AC \mQ\); furthermore
	finite \(\RMI{\rno}{\mP}{\Wm}\) hypothesis of Lemma \ref{lem:Augustinmeanexistence} 
	implies \(\qmn{\widetilde{\mP}}\!=\!\qmn{\mP}\). 
	Thus \(\qmn{\widetilde{\mP}}\AC \mQ\) hypothesis can be omitted
	for \(\rno\!\in\!(1,\infty)\).
	For \(\rno\!\in\!(0,1)\), however, \(\qmn{\widetilde{\mP}}\AC \mQ\) hypothesis  cannot be dropped;
	see \cite[footnote 11]{nakiboglu19C}, and \cite[(15)]{chengHT19}, \cite[Thm.~IV.14]{mosonyiO21} 
	for classical-quantum channels, and a related problem in \cite[Lem.~5]{HayashiTomamichel14}.
\end{remark}

\begin{proof}
	For \(\rno\!=\!1\) case lemma follows from 
	\eqref{eq:pinkser} and \eqref{eq:sibson-one}
	for \(\qmn{1,\mP}\!=\!\qmn{\mP}\).
	For other orders, first apply Lemma \ref{lem:monotonicity} 
	for \(\rnb=1\wedge \sfrac{1}{\rno}\)\vspace{-.1cm}
	\begin{align}
	\notag
	\CRD{\rno}{\Wm}{\mQ}{\mP}
	\!-\!
	\CRD{\rno}{\Wm}{\tAop{\rno}{\mP}{1\wedge \sfrac{1}{\rno}}{\mQ}}{\mP}
	&\!\geq\!\tfrac{1\wedge\rno}{2\rno}\lon{\Aop{\rno}{\mP}{\mQ}\!-\!\mQ}^2.
	\end{align}
	Then \eqref{eq:lem:fixedpoint} follows from by \eqref{eq:def:augustininformation} and \eqref{eq:lem:Augustinmeanexistence}.

	For \(\rno\!\in\!(0,1)\), \(\Wma{\rno}{\mQ}(\dinp)\AC\mQ\) 
	whenever \(\Wma{\rno}{\mQ}(\dinp)\) is defined.
	Thus using \eqref{eq:def:Aoperator}, \eqref{eq:lem:fixedpoint},
	and  \(\qmn{\rno,\mP}\AC\qmn{\mP}\) established in Lemma \ref{lem:Augustinmeanexistence}, 
	we can obtain the following identity for all \(\rno\in(0,1)\),\vspace{-.1cm}
	\begin{align}
	\label{eq:meannew}
	\der{\qmn{\rno,\mP}}{\qmn{\mP}}
	&\!=\!\left(\EXS{\mP\!}{\left(\der{\Wmn{\sim}(\inp)}{\qmn{\mP}}\right)^{\rno} e^{(1-\rno)\RD{\rno}{\Wm(\inp)}{\qmn{\rno,\mP}}}}\right)^{\sfrac{1}{\rno}},
	&
	&
	\\
	\notag
	&\!\geq\!\left(\EXS{\mP\!}{\left(\der{\Wmn{\sim}(\inp)}{\qmn{\mP}}\right)^{\rno}}\right)^{\sfrac{1}{\rno}},
	&
	&
\end{align}~\vspace{-.45cm}\\
where the inequality follows from \(\RD{\rno}{\Wm(\inp)}{\qmn{\rno,\mP}}\!\geq\!0\). 	
Thus \(\qmn{\widetilde{\mP}}\AC\qmn{\rno,\mP}\) by \eqref{eq:def:outputdistributionAC} because 
\(\EX{\sta^{\rno}}\!>\!0\) iff \(\EX{\sta}\!>\!0\) for any non-negative random variable \(\sta\).
Furthermore, for any  \(\rno\!\in\!(0,1)\)\vspace{-.17cm}
\begin{align}
\notag
\CRD{\rno}{\Wm}{\mQ}{\mP}
&\stackrel{(a)}{=}\! 
\CRD{\rno}{\Wm}{\qmn{\sim}}{\mP}
\\
\notag
&\stackrel{(b)}{=}\! 
\tfrac{1}{\rno-1}\EXS{\mP\!}{\!\ln\!\!\int\!\!
	\left(\der{\Wmn{\!\sim\!}(\inp)}{\qmn{\mP}}\right)^{\rno}\!\left(\der{\qmn{\sim}}{\qmn{\mP}}\right)^{1-\rno}\!\!
	\qmn{\mP}(\dif{\dout})\!}
\\
\notag
&\stackrel{(c)}{=}\!
\tfrac{1}{\rno-1}\EXS{\mP\!}{\!\ln\!\!\int\!\!
	\left(\der{\Wmn{\!\sim\!}(\inp)}{\qmn{\mP}}\right)^{\rno}\!\left(\der{\qmn{ac}}{\qmn{\mP}}\right)^{1-\rno}\!\!
	\qmn{\mP}(\dif{\dout})\!}
\\
\label{eq:absolutelycontinuouspart2}
&\stackrel{(d)}{=}\CRD{\rno}{\Wm}{\qmn{ac}}{\mP},
\end{align}~\vspace{-.45cm}\\
	where 
	\(\qmn{\sim}\) is \(\qmn{\mP}\)-absolutely continuous part of \(\mQ\),
	\(\qmn{ac}\) is \(\qmn{\widetilde{\mP}}\)-absolutely continuous part 
	of both \(\qmn{\sim}\) and  \(\mQ\),
	\((a)\) follows from \eqref{eq:absolutelycontinuouspart},
	\((b)\) follows from the definition of \renyi divergence for \(\rno\!\in\!(0,1)\)
	and \(\qmn{\sim}\AC\qmn{\mP}\),
	\((c)\) follows from \eqref{eq:def:outputdistributionAC}
	because as a result only the \(\qmn{\widetilde{\mP}}\)-absolutely continuous 
	part of \(\qmn{\sim}\) contributes to the integral \(\mP\)-a.s.,
	\((d)\) follows from the definition of \renyi divergence of \(\rno\!\in\!(0,1)\)
	and \(\qmn{ac}\AC\qmn{\mP}\) .
	Note that \eqref{eq:absolutelycontinuouspart2} implies 
	\(\qmn{\rno,\mP}\AC\qmn{\widetilde{\mP}}\)
	and hence \(\qmn{\rno,\mP}\sim\qmn{\widetilde{\mP}}\) for \(\rno\in(0,1)\)
	because we have already established \(\qmn{\widetilde{\mP}}\AC\qmn{\rno,\mP}\).
	
	For \(\rno\!\in\!(1,\infty)\),
	\(\CRD{\rno}{\Wm}{\qmn{\rno,\mP}}{\mP}\!<\!\infty\)
	implies \(\Wm(\dinp)\AC\qmn{\rno,\mP}\) \(\mP\)-a.s.~and \(\qmn{\widetilde{\mP}}\!=\!\qmn{\mP}\).
	Thus \(\qmn{\mP}\AC \qmn{\rno,\mP}\) by \eqref{eq:outputdistribution}
	and consequently \(\qmn{\mP}\sim\qmn{\rno,\mP}\) 
	by Lemma \ref{lem:Augustinmeanexistence}.
	Thus \(\qmn{\rno,\mP}\sim\qmn{\widetilde{\mP}}\),
	for \(\rno\!\in\!(1,\infty)\).

Let \(\mS\in\pmea{\outA}\) satisfy \(\Aop{\rno}{\mP}{\mS}=\mS\)  and \(\qmn{\widetilde{\mP}}\AC \mS\),
and \(\qmn{ac}\) be \(\qmn{\widetilde{\mP}}\)-absolutely continuous part of a \(\mQ\in\pmea{\outA}\).
For \(\rno>1\), finite \(\RMI{\rno}{\mP}{\Wm}\) hypothesis implies 
\(\qmn{\widetilde{\mP}}=\qmn{\mP}\).
Then invoking \eqref{eq:absolutelycontinuouspart2} for \(\rno\in(0,1)\)
and \eqref{eq:absolutelycontinuouspart} for \(\rno\in(1,\infty)\) we get\vspace{-.15cm}
	\begin{align}
	\notag
	\hspace{1cm}&\hspace{-1cm}
	\CRD{\rno}{\Wm}{\mQ}{\mP}
	\!-\!
	\CRD{\rno}{\Wm}{\mS}{\mP}
	\\
	\notag
	&=\tfrac{1}{\rno-1}\EXS{\mP\!}{\ln \int  
		\left(\der{\qmn{ac}}{\mS}\right)^{1-\rno}\Wma{\rno}{\mS}(\dif{\dout}|\inp)}
	\\
	\notag
	&\stackrel{(a)}{\geq} 
	\begin{cases}
	\tfrac{1}{\rno-1}\ln \EXS{\mP\!}{\displaystyle{\int}
		\left(\der{\qmn{ac}}{\mS}\right)^{1-\rno}\Wma{\rno}{\mS}(\dif{\dout}|\inp)}
	&\text{if~}\rno\!<\!1
	\\
	\tfrac{1}{\rno-1}  \EXS{\mP\!}{\displaystyle{\int}
		\ln \left(\der{\qmn{ac}}{\mS}\right)^{1-\rno}\!\Wma{\rno}{\mS}(\dif{\dout}|\inp)}
	&\text{if~}\rno\!>\!1
	\end{cases}
	\\
	\notag
	&\stackrel{(b)}{=}
	\begin{cases}
	\tfrac{1}{\rno-1}\ln \displaystyle{\int}
	\der{\Aop{\rno}{\mP}{\mS}}{\mS}\left(\der{\qmn{ac}}{\mS}\right)^{1-\rno}
	\mS(\dif{\dout})
	&\text{if~}\rno\!<\!1
	\\
	-\displaystyle{\int}\der{\Aop{\rno}{\mP}{\mS}}{\mS}\ln \left(\der{\qmn{ac}}{\mS}\right) \mS(\dif{\dout})
	&\text{if~}\rno\!>\!1
	\end{cases}
	\\
	\label{eq:lowerboundonthedifference}
	&\stackrel{(c)}{\geq}
	\RD{1\wedge\rno}{\mS}{\mQ}
	\end{align}
where \((a)\) follows from Jensen's inequality and
the concavity of natural logarithm function,
\((b)\)  follows from \eqref{eq:def:Aoperator}
and Fubini's theorem \cite[Thm. 3.4.4]{bogachev},
\((c)\) follows from \(\Aop{\rno}{\mP}{\mS}\!=\!\mS\),
\cite[Lemma 1]{nakiboglu19C}, and \(\qmn{ac}\!\leq\!\mQ\).
Thus
\(\CRD{\rno}{\Wm}{\mQ}{\mP}\!>\!\CRD{\rno}{\Wm}{\mS}{\mP}\)
for all \(\mQ\!\in\!\pmea{\outA}\setminus\{\mS\}\) by
\eqref{eq:pinkser}
and \(\mS=\qmn{\rno,\mP}\) by Lemma \ref{lem:Augustinmeanexistence},
for any \(\mS\!\in\!\pmea{\outA}\) satisfying both 
\(\Aop{\rno}{\mP}{\mS}\!=\!\mS\)  and \(\qmn{\widetilde{\mP}}\AC\mS\).
Proof of \eqref{eq:lem:EHB}  
is presented in Appendix \ref{sec:FixedPointProofPart}.	
\end{proof}
\begin{remark}
The identity \eqref{eq:meannew} holds not only for \(\rno\!\in\!(0,1)\) but 
for any \(\rno\in\reals{+}\) satisfying \(\RMI{\rno}{\mP}{\Wm}\!<\!\infty\). 
For \(\rno\!\in\!(1,\infty)\) case, if \(\RMI{\rno}{\mP}{\Wm}\!<\!\infty\)
then  \eqref{eq:meannew} follows from 
\eqref{eq:def:Aoperator}, \eqref{eq:lem:fixedpoint}, \(\qmn{\rno,\mP}\AC\qmn{\mP}\),
and the fact that \(\Wm(\dinp)\AC\qmn{\rno,\mP}\) \(\mP\)-a.s.~and it can be written 
as\vspace{-.15cm}
\begin{align}
\label{eq:mean}
\der{\qmn{\rno,\mP}}{\qmn{\mP}}
&=\left(\EXS{\mP\!}{\left(\der{\Wm(\inp)}{\qmn{\mP}}\right)^{\rno} e^{(1-\rno)\RD{\rno}{\Wm(\inp)}{\qmn{\rno,\mP}}}}\right)^{\sfrac{1}{\rno}}.
&
&
\end{align}~\vspace{-.45cm}\\
For \(\rno\in(0,1)\) case,  \eqref{eq:mean} holds whenever \(\Wm(\dinp)\AC\qmn{\mP}\) \(\mP\)-a.s., 
e.g. when \(\mP\) is a probability mass function as in \cite[(38)]{nakiboglu19C}. 
\end{remark}
\begin{remark}\label{rem:definingAugustinMeanAsTheFixedPoint}
	In \cite{verdu21}, the channel \(\Wm\) is assumed to satisfy \(\Wm(\dinp)\AC\qmn{\mP}\) \(\mP\)-a.s.~for all \(\mP\), which is a reasonable but not completely general assumption. 
	Ref.~\cite{verdu21} defines the Augustin mean, which it calls \(\langle\rno\rangle\)-response to \(\mP\),
	as the element of \(\pmea{\qmn{\mP}}\) satisfying \eqref{eq:mean}; see \cite[(92)]{verdu21}.
	The existence of a unique element of \(\pmea{\qmn{\mP}}\) satisfying \eqref{eq:mean}, however, is 
	not a definition, but an assertion that requires a proof. 
	Furthermore, the proof of Lemma 
	{\href{https://arxiv.org/pdf/1803.07937.pdf#Item.44}{13-(c)-i}} and 
	{\href{https://arxiv.org/pdf/1803.07937.pdf#Item.52}{13-(d)-i}}
	in \cite{nakiboglu19C},
	had previously shown 
for any probability mass function \(\mP\) with a finite support set that 
when a \(\mQ\in\pmea{\outA}\) satisfying both 
\(\Aop{\rno}{\mP}{\mQ}\!=\!\mQ\) and \(\qmn{\mP}\AC\mQ\) exists, 
it has to be the Augustin mean,
and these arguments are valid as they are for general input distributions \(\mP\), as well.
\end{remark}
\begin{example}[A Channel-Input Distribution Pair for which \(\Wm(\dinp)\NAC\qmn{\mP}\) \(\mP\)-a.s.]\label{eg:partiallynoiseless}
Let the probability density function of 
the channel output \(\dout\in (0,2)\) 
given the channel input \(\dinp\in (0,1)\),
\(\mW(\dout|\dinp)\) be
\begin{align}
\label{eq:partiallynoiselesschannel}
\mW(\dout|\dinp)
&=\tfrac{\IND{\dout\in(0,\dinp)}+(\dout-\dinp)\IND{\dout\in(0,1)}
+(\gamma-0.5)\delta(\dout-\dinp-1)}{\gamma}, \vspace{-1em}
\end{align}
where \(\IND{\cdot}\) is the indicator function,
\(\delta(\cdot)\) is the Dirac delta function,
and \(\gamma\) is a constant in \((0.5,\infty)\).
Let the input distribution \(\mP\) be the uniform distribution
on \((0,1)\) then the Radon--Nikodym derivarives
of \(\qmn{\mP}\) and \(\qmn{\widetilde{\mP}}\) 
with respect to the Lebesgue measure are\vspace{-.1cm}
\begin{align}
\notag
\der{\qmn{\mP}}{\lambda}
&=\tfrac{\IND{\dout\in(0,1)}+(2\gamma-1)\IND{\dout\in(1,2)}}{2\gamma}, \vspace{-1em}
\\
\notag
\der{\qmn{\widetilde{\mP}}}{\lambda}
&=\tfrac{\IND{\dout\in(0,1)}}{2\gamma}.
\end{align}
Note that \(\Wm(\dinp)\NAC\qmn{\mP}\) for all \(\dinp\in(0,1)\).
Nevertheless, the Augustin information can be calculated
for all positive orders:\vspace{-.1cm}
\begin{align}
	\RMI{\rno}{\mP}{\Wm}
	&=\begin{cases}
		\tfrac{\rno \ln \gamma+ \ln(1+\rno)}{1-\rno}
		&\text{if~}\rno\in(0,1)
		\\
		\infty
		&\text{if~}\rno\in[1,\infty)
	\end{cases}.
\end{align}
Furthermore, for all \(\rno\in(0,1)\), 
the order \(\rno\) Augustin mean is the uniform distribution on \((0,1)\)
and \eqref{eq:meannew} holds for all \(\rno\!\in\!(0,1)\),
as expected.
\end{example}
\section{Discussions} \label{sec:discussions}
Augustin information was defined for arbitrary channels with 
countably generated output \(\sigma\)-algebras in \cite[\S5.4]{nakiboglu19C}.
The existence of a unique Agustin center was confirmed both for
the unconstrained and cost constrained cases for
channels with countably separated input \(\sigma\)-algebras,
provided that Augustin capacity is finite;
see \cite[Thms 4 and 5]{nakiboglu19C}.
However, the existence of a unique Augustin mean was not proved 
for general input distributions on these channels in \cite{nakiboglu19C}.

The technical challenge arises from the lack of closed form expression
for the minimizer in \eqref{eq:def:augustininformation}.
If the output set is finite, then the probability simplex is compact; thus 
the lower semicontinuity and the extreme value theorem 
implies the existence of a minimizer.
When the output space is an arbitrary measurable space, 
the existence of the minimizer has only been proved 
for input distributions with finite support set,
\cite{augustin78,nakiboglu19C,chengGH19}.
In these proofs, finite support of the input distribution is
used to reach an intermediary problem with compactness.  
Thus previous proofs of the existence of the Augustin mean 
relied on some form of compactness directly.

The novelty of our approach is the use of 
Banach--Saks property and convexity in lieu of compactness. 
We lift the optimization in \eqref{eq:def:augustininformation} 
from the set of all probability measures 
to an \(L^{\tau}\) space for a \(\tau\!>\!1\)
because 
the space of probability measure \({\cal P}\) does not have 
the Banach--Saks 
property.\footnote{One might think of working in \(L^1\) instead of \(L^{\tau}\) and
	invoking the Koml\'os theorem 
	\cite[Theorem 1.a]{komlos67},\cite[4.7.24 Theorem]{bogachev}---every 
	norm bounded sequence in \(L^1\) 
	contains a subsequence whose Ces\`aro mean converges almost everywhere. 
	However, this fact alone does not guarantee the setwise convergence 
	that is crucial to the application of lower semicontinuity of 
	the \renyi divergence in its second argument.}
Despite the change in the underlying vector space structure,
the new functional \(\hhx{\rno}{\cdot}\) 
inherits both the convexity and norm lower semicontinuity 
from the \renyi divergence, for an appropriately chosen \(\tau\). 
Use of the Banach--Saks theorem in conjunction with 
the (quasi-)convexity and the norm lower semicontinuity 
of the objective function to prove the existence of its minimizer 
seems to be a novel approach  more generally in the context of 
information theoretic optimization problems.

\appendix
\subsection{Proof of Lemma \ref{lem:inherited_convexity}} \label{sec:convexity}
	Note that for \(\rno\geq 1\),
	if there exists an \(\fX\in\Bset{}{\tau_{\rno}}{\qmn{\mP}}\)	
	such that \(\hhx{\rno}{\fX}\!<\!\infty\),
	then \(\hhx{1}{\fX}\!<\!\infty\) and
	\(\Wm(\dinp)\AC\qmn{\mP}\) \(\mP\)-a.s.
	For \(\rno\!\in\!(0,1)\),
	if there exists an \(\fX\in\Bset{}{\tau_{\rno}}{\qmn{\mP}}\)	
	such that \(\hhx{\rno}{\fX}\!<\!\infty\),
	then \(\lon{\Wmn{\sim}(\dinp)}\!>\!0\) \(\mP\)-a.s.,
	where \(\Wmn{\sim\!}(\dinp)\) is the
	\(\qmn{\mP}\)-absolutely continuous component of \(\Wm(\dinp)\) 
	for all \(\dinp\!\in\!\inpS\).
	In either case, \(\hhx{\rno}{\fX}\) can be expressed 
	in terms of \(\Wmn{\sim\!}\) as follows\vspace{-.1cm}
	\begin{align}
	\label{eq:inherited_convexity:1}
	\hhx{\rno}{\fX} 
	&=\EXS{\mP\!}{\tfrac{1}{\rno-1}\ln \EXS{\qmn{\mP}}{\hX_{\inp}^{\rno}\fX^{\tau_\rno(1-\rno)} }},
	\end{align}	
	where \(\hX_{\dinp}\!\DEF\!\der{ \Wmn{\sim}(\dinp)}{\qmn{\mP}}\)
	for all \(\dinp\!\in\!\inpS\).
	
	We establish the convexity of \(\hhx{\rno}{\cdot}\) by invoking 
	\eqref{eq:inherited_convexity:1}, but we need to modify other 
	ingredients of the proof based on the value of \(\rno\).
	Let us first consider \(\rno\in (0,0.5)\) case:\vspace{-.1cm}
	\begin{align}
	\notag
	\hspace{1cm}&\hspace{-1cm}
	\hhx{\rno}{\rnb\fX+(1-\rnb)\gX} 
	\\
	\notag
	&=
	\EXS{\mP\!}{
		\tfrac{1}{\rno-1}\ln 
		\EXS{\qmn{\mP}}{\hX_{\inp}^{\rno}(\rnb\fX+(1-\rnb)\gX)}
	}
	\\
	\notag
	&=
	\EXS{\mP\!}{
		\tfrac{1}{\rno-1}\ln 
		\left(\rnb\EXS{\qmn{\mP}}{\hX_{\inp}^{\rno}  \fX}
		+ (1 - \rnb)\EXS{\qmn{\mP}}{\hX_{\inp}^{\rno}  \gX}
		\right)   
	}
	\\
	\notag
	&\stackrel{(a)}{\leq} 
	\EXS{\mP\!}{
		\tfrac{\rnb}{\rno-1}\ln 
		\EXS{\qmn{\mP}}{\hX_{\inp}^{\rno}\fX}
		+\tfrac{1-\rnb}{\rno-1}\ln 
		\EXS{\qmn{\mP}}{\hX_{\inp}^{\rno}\gX}
	}
	\\
	\notag
	&=\rnb\hhx{\rno}{\fX}+(1-\rnb)\hhx{\rno}{\gX},
	\end{align}
	where \((a)\) follows from Jensen's inequality and the concavity of the natural logarithm function.
	
	Next, we move onto the case \(\rno\in [0.5,1)\):\vspace{-.1cm}
	\begin{align}
	\notag
	\hspace{0.15cm}&\hspace{-0.15cm}
	\hhx{\rno}{\rnb\fX+(1-\rnb)\gX} \\
	&\!=\!
	\EXS{\mP\!}{
		\tfrac{1}{\rno-1}\ln 
		\EXS{\qmn{\mP}}{\hX_{\inp}^{\rno}(\rnb\fX\!+\!(1\!-\!\rnb)\gX)^{2(1-\rno)}}
	} \notag
	\\ \notag
	&\!\stackrel{(a)}{\leq}\!
	\EXS{\mP\!}{
		\tfrac{1}{\rno-1}\ln 
		\EXS{\qmn{\mP}}{\hX_{\inp}^{\rno}(\rnb\fX^{2(1-\rno)}
			\!+\!(1\!-\!\rnb)\gX^{2(1-\rno)})}
	} 
	\\ \notag
	&\!=\!\EXS{\mP\!}{\tfrac{1}{\rno-1}\ln 
		\EXS{\qmn{\mP}}{\rnb\hX_{\inp}^{\rno}\fX^{2(1-\rno)}\!+\!(1\!-\!\rnb)\hX_{\inp}^{\rno}\gX^{2(1-\rno)}}}
	\\\notag
	&\!\stackrel{(b)}{\leq}\! 
	\EXS{\mP\!}{
		\tfrac{\rnb}{\rno-1}\ln 
		\EXS{\qmn{\mP}\!}{\hX_{\inp}^{\rno}\fX^{2(1-\rno)}}
		\!+\!\tfrac{1-\rnb}{\rno-1}\ln 
		\EXS{\qmn{\mP}\!	}{\hX_{\inp}^{\rno}\gX^{2(1-\rno)}}
	}
	\\
	\notag
	&=\rnb\hhx{\rno}{\fX}+(1-\rnb)\hhx{\rno}{\gX},
	\end{align}
	where \((a)\) follows from Jensen's inequality and the concavity of the power function \((\cdot)^{2(1-\rno)}\), and \((b)\) follows from Jensen's inequality and the concavity of the natural logarithm function.
	
	For \(\rno\in[1,\infty]\) case,  first note that \eqref{eq:def:inverseRND} implies\vspace{-.1cm} 
	\begin{align}
	\notag 
	\hqx{2}{\rnb\fX+(1-\rnb)\gX} 
	&=\rnb^2\hqx{2}{\fX}+2\rnb(1-\rnb)\hqx{2}{\sqrt{\fX\gX}}\hspace{.3cm}
	\\
	\notag 
	&\hspace{1cm}+(1-\rnb)^2\hqx{2}{\gX}.
	\end{align}
	Then the convexity of \renyi divergence, \cite[Thm. 12]{ervenH14} implies\vspace{-.1cm}
	\begin{align}
	\notag
	\hspace{0.2cm}&\hspace{-0.2cm}
	\hhx{\rno}{\rnb\fX+(1-\rnb)\gX} \\
	\notag
	&\!\leq\! 
	\rnb^{2}\hhx{\rno}{\fX}
	\!+\!2\rnb(1\!-\!\rnb)\hhx{\rno}{\sqrt{\fX\gX}}
	\!+\!(1\!-\!\rnb)^{2}\hhx{\rno}{\gX}
	\\	
	\notag 
	&\!\stackrel{(a)}{\leq}\! 
	\rnb^{2}\hhx{\rno}{\fX}
	\!+\!\rnb(1\!-\!\rnb) (\hhx{\rno}{\fX}\!+\!\hhx{\rno}{\gX})
	\!+\!(1\!-\!\rnb)^{2}\hhx{\rno}{\gX}  
	\\
	\notag
	&\!=\!\rnb\hhx{\rno}{\fX}+(1-\rnb)\hhx{\rno}{\gX},
	\end{align}
	where \((a)\) follows from inequality \(\hhx{\rno}{\sqrt{\fX\gX}} \!\leq\! \tfrac{\hhx{\rno}{\fX}+\hhx{\rno}{\gX}}{2}\) 
	established for different values of \(\rno\) in
	\eqref{eq:hqrt-one}, \eqref{eq:hqrt-oti},  and \eqref{eq:hqrt-inf}.\vspace{-.1cm}
	\begin{align}
	\notag
	\hhx{1}{\sqrt{\fX\gX}}
	&=\EXS{\mP\!}{\EXS{\qmn{\mP}}{\hX_{\inp}
			\ln \tfrac{\hX_{\inp}}{\fX\gX}}}
	\hspace{4.5cm} \\
	\label{eq:hqrt-one}
	&=\tfrac{\hhx{1}{\fX}+\hhx{1}{\gX}}{2}.
	\end{align}
	For \(\rno\in(1,\infty)\),\vspace{-.1cm}
	\begin{align}
	\notag
	\hspace{.5cm}&\hspace{-.5cm}
	\hhx{\rno}{\sqrt{\fX\gX}} \\
	\notag
	&\!=\!\tfrac{1}{\rno-1}\EXS{\mP\!}{\ln 
		\EXS{\qmn{\mP}}{\hX_{\inp}^{\rno}(\fX\gX)^{1-\rno}}}
	\\
	\notag
	&\!\stackrel{(a)}{\leq}\!\tfrac{1}{\rno-1}  
	\EXS{\mP\!}{\ln
		\sqrt{\EXS{\qmn{\mP}}{\hX_{\inp}^{\rno}\fX^{2(1-\rno)}}
			\EXS{\qmn{\mP}}{\hX_{\inp}^{\rno}\gX^{2(1-\rno)}}}
	}
	\\
	\notag
	&\!=\!\tfrac{1}{2}\tfrac{1}{\rno-1}
	\EXS{\mP\!}{
		\ln \EXS{\qmn{\mP}}{\hX_{\inp}^{\rno}\fX^{2(1-\rno)}}
		\!+\!\ln \EXS{\qmn{\mP}}{\hX_{\inp}^{\rno}\gX^{2(1-\rno)}}
	}
	\\
	\label{eq:hqrt-oti}
	&\!=\!\tfrac{\hhx{\rno}{\fX}+\hhx{\rno}{\gX}}{2},
	\end{align}
	where \((a)\) follows from the Cauchy--Schwarz inequality.
	\begin{align}\vspace{-.1cm}
	\notag
	\hhx{\infty}{\sqrt{\fX\gX}} 
	&=\EXS{\mP\!}{\ln \essup\nolimits \tfrac{\hX_{\inp}}{\fX\gX}}
	\\
	\notag
	&\leq
	\EXS{\mP\!}{
		\tfrac{1}{2}\ln\essup\tfrac{\hX_{\inp}}{\fX^2}
		+
		\tfrac{1}{2}\ln\essup\tfrac{\hX_{\inp}}{\gX^2}
	}
	\\
	\label{eq:hqrt-inf}
	&=\tfrac{\hhx{\infty}{\fX}+\hhx{\infty}{\gX}}{2}.
	\end{align}

\subsection{Proof of Lemma \ref{lem:inherited_lsc}}\label{sec:lsc}
The norm lower semicontinuity of the functional 
\(\hhx{\rno}{\cdot}\) follows from
the norm continuity of the function
\(\hqx{\tau_{\rno}\!}{\cdot}\)
for the total variation topology on its range
and the norm lower semicontinuity of 
\(\RD{\rno}{\mW}{\cdot}\) on \(\fmea{\outA}\).

Let us start with establishing 
the continuity of \(\hqx{\tau_{\rno}}{\cdot}\).
Note that for all \(\tau \in (1,2]\) and
\(\mA,\mB\in\reals{\geq 0}\) we have,\vspace{-.1cm}
\begin{align}
\notag 
(\mA^{\tau}\!-\!\mB^{\tau})
\!+\!\mA^{\tau-1}\mB^{\tau-1}
(\mA^{2-\tau}\!-\!\mB^{2-\tau})
&\!=\!
(\mA\!-\!\mB)(\mA^{\tau-1}\!+\!\mB^{\tau-1}).
\end{align}
Furthermore, 
\((\mA^{\tau}\!-\!\mB^{\tau})\)
and
\((\mA^{2-\tau}\!-\!\mB^{2-\tau})\)
never have opposite signs.
Thus for all \(\tau \in (1,2]\) and
\(\mA,\mB\in\reals{\geq 0}\) we have\vspace{-.1cm}
\begin{align}
\notag
\abs{\mA^{\tau}-\mB^{\tau}}
&\leq 
\abs{\mA-\mB}(\mA^{\tau-1}+\mB^{\tau-1}).
\end{align}
Then\vspace{-.1cm}
\begin{align}
\notag \EXS{\qmn{\mP}}{\abs{\fX^{\tau}\!-\!\gX^{\tau}}}
&\!\leq\! 
\EXS{\qmn{\mP}}{\abs{\fX-\gX}(\fX^{\tau-1}+\gX^{\tau-1})}
\\
\notag
&\!\stackrel{(a)}{\leq} \!
\EXS{\qmn{\mP}}{\abs{\fX\!-\!\gX}^{\tau}}^{\frac{1}{\tau}}
\EXS{\qmn{\mP}}{(\fX^{\tau-1}\!+\!\gX^{\tau-1})^{\frac{\tau}{\tau-1}}}^{\frac{\tau-1}{\tau}} 
\\
\notag
&\!=\!
\nrm{\tau}{\fX\!-\!\gX}
2\EXS{\qmn{\mP}}{\left(\tfrac{\fX^{\tau-1}}{2}\!+\!\tfrac{\gX^{\tau-1}}{2}\right)^{\frac{\tau}{\tau-1}}}^{\frac{\tau-1}{\tau}}
\\
\notag
&\!\stackrel{(b)}{\leq}\!  
\nrm{\tau}{\fX\!-\!\gX}
2\EXS{\qmn{\mP}}{\left(\tfrac{\fX}{2}\!+\!\tfrac{\gX}{2}\right)^{\tau}}^{\frac{\tau-1}{\tau}} 
\\
\notag
&\!=\!\nrm{\tau}{\fX\!-\!\gX}
2^{2-\tau}
\nrm{\tau}{\fX\!+\!\gX}^{\tau-1}
\\	
\notag
&\stackrel{(c)}{\leq} 
2^{2-\tau}
\nrm{\tau}{\fX-\gX}
\left(2\nrm{\tau}{\fX}+\nrm{\tau}{\fX-\gX}
\right)^{\tau-1}, 
\end{align}
where 
\((a)\) follows from H\"older's inequality; 
\((b)\) follows from Jensen's inequality and the concavity of the power function \((\cdot)^{\tau-1}\)
for \(\tau\!\in\!(1,2]\), and 
\((c)\) follows from the triangle inequality for the \(\tau\)-norm.
Then the continuity of \(\hqx{\tau_{\rno}}{\cdot}\) follows from
the identity \(\nrm{}{\hqx{\tau}{\fX}-\hqx{\tau}{\gX}}= \notag \EXS{\qmn{\mP}}{\abs{\fX^{\tau}-\gX^{\tau}}}\) 
and the fact that \(\tau_{\rno} \in (1,2]\).

Now we are left with establishing the norm lower semicontinuity 
of \(\RD{\rno}{\mW}{\cdot}\) on \(\fmea{\outA}\).
To that end first note that\vspace{-.1cm} 
\begin{align}
\notag
\RD{\rno}{\mW}{\mQ}
&=\RD{\rno}{\mW}{\sfrac{\mQ}{\lon{\mQ}}}-\ln \lon{\mQ}
&
&\forall \mQ:\lon{\mQ}>0.
\end{align}
Then \(\RD{\rno}{\mW}{\cdot}\) is continuous at the zero measure by 
\eqref{eq:pinkser} because \(\RD{\rno}{\mW}{0}=\infty\).
On the other hand, for non-zero measures 
\(\RD{\rno}{\mW}{\sfrac{\cdot}{\lon{\cdot}}}\) 
is norm lower semicontinuous on \(\fmea{\outA}\)
because \(\RD{\rno}{\mW}{\cdot}\) is
lower semicontinuous on \(\pmea{\outA}\) for the topology of
setwise convergence by \cite[Thm. 15]{ervenH14}
and \(\sfrac{\cdot}{\lon{\cdot}}\) 
is norm continuous for the topology of setwise convergence 
on its range \(\pmea{\outA}\).
Hence, \(\RD{\rno}{\mW}{\cdot}\) is 
norm lower semicontinuous on \(\fmea{\outA}\)
for non-zero measures, as well, as a result  
of the  continuity of the natural logarithm function.

\subsection{Proof of Lemma \ref{lem:Augustinmeanexistence}}\label{sec:ExitenceProof}
There exists an \(\fX_{\rno}\!\in\!\Lp{\tau_{\rno}}{\qmn{\mP}}\) satisfying 
both \(\hqx{\tau_{\rno}}{\fX_{\rno}}\!\in\!\pmea{\outA}\) and
\(\CRD{\rno}{\Wm}{\hqx{\tau_{\rno}}{\fX_{\rno}}}{\mP}\!=\!\RMI{\rno}{\mP}{\Wm}\)
by Lemma \ref{lem:existence}.
Furthermore, \(\qmn{\rno,\mP}\AC \qmn{\mP}\) by 
the definition of \(\hqx{\tau}{\fX_{\rno}}\) given in 
\eqref{eq:def:inverseRND}. 

To establish that \(\hqx{\tau_{\rno}}{\fX_{\rno}}\)
is the only probability measure
achieving the infimum in \eqref{eq:def:augustininformation},
first note that 
\eqref{eq:def:augustininformation} and
\eqref{eq:absolutelycontinuouspart} imply
\begin{align}
\notag
\RMI{\rno}{\mP}{\Wm}
&<\CRD{\rno}{\Wm}{\mQ}{\mP}
&
&\forall \mQ\in\pmea{\outA}\setminus\pmea{\qmn{\mP}}.
\end{align}
That is \(\RMI{\rno}{\mP}{\Wm}=\CRD{\rno}{\Wm}{\mQ}{\mP}\)
can hold only for \(\mQ\)'s in \(\pmea{\outA}\) that are absolutely continuous in \(\qmn{\mP}\).
On the other hand, for any \(\dinp\in\inpS\),
\(\smn{1},\smn{0}\in\pmea{\qmn{\mP}}\),
and \(\rnb\!\in\!(0,1)\),
the strict convexity of the \renyi divergence described in \cite[Thm. 12]{ervenH14}
implies
\begin{align}
\notag
\RD{\rno}{\Wm(\dinp)}{\smn{\rnb}}
&\leq 
\rnb\RD{\rno}{\Wm(\dinp)}{\smn{1}}
+(1-\rnb)\RD{\rno}{\Wm(\dinp)}{\smn{0}},
\end{align}
for \(\smn{\rnb}\!=\!\rnb\smn{1}\!+\!(1\!-\!\rnb)\smn{0}\)
and the equality holds iff
\(\der{\smn{0}}{\qmn{\mP}}=\der{\smn{1}}{\qmn{\mP}}\) holds \(\Wm(\dinp)\)-a.s.
Thus, for any \(\smn{1},\smn{0}\in\pmea{\qmn{\mP}}\) and \(\rnb\!\in\!(0,1)\)
\begin{align}
\notag
\CRD{\rno}{\Wm}{\smn{\rnb}}{\mP}
&\leq
\rnb\CRD{\rno}{\Wm}{\smn{1}}{\mP}
+(1-\rnb)\CRD{\rno}{\Wm}{\smn{0}}{\mP},
\end{align}
and the equality holds iff \(\mP({\cal S}_{\smn{0},\smn{1}})\!=\!0\), where
\({\cal S}_{\smn{0},\smn{1}}\in\inpA\) and \(\oev_{\smn{0},\smn{1}}\!\in\!\outA\) 
are defined as follows
\begin{align}
\notag	
\oev_{\smn{0},\smn{1}}
&\DEF\left\{\dout: \der{\smn{0}}{\qmn{\mP}}\neq \der{\smn{1}}{\qmn{\mP}}\right\},
\\
\notag	
{\cal S}_{\smn{0},\smn{1}}
&\DEF\left\{\dinp: \Wm(\oev_{\smn{0},\smn{1}}|\dinp)>0\right\}.
\end{align}
But \(\qmn{\mP}(\oev_{\smn{0},\smn{1}})\!>\!0\) for any \(\smn{0}\neq\smn{1}\).
Thus \(\mP({\cal S}_{\smn{0},\smn{1}})\!>\!0\) for any \(\smn{0}\neq\smn{1}\)
by \eqref{eq:outputdistribution}.
Consequently, the infimum in \eqref{eq:def:augustininformation}
cannot be achieved by two distinct elements of \(\pmea{\qmn{\mP}}\),
either.
Hence, \(\qmn{\rno,\mP}\!=\!\hqx{\tau_{\rno}}{\fX_{\rno}}\)
is the only distribution in \(\pmea{\outA}\)
achieving the infimum in \eqref{eq:def:augustininformation}.

\subsection{Proof of \eqref{eq:lem:EHB} of Lemma \ref{lem:fixedpoint}}\label{sec:FixedPointProofPart}
The lower bound given in \eqref{eq:lem:EHB} for the difference  
follows from \eqref{eq:lowerboundonthedifference} for \(\mS\!=\!\qmn{\rno,\mP}\).
To prove the upper bound given in \eqref{eq:lem:EHB},
for the difference, let us denote
\(\qmn{\widetilde{\mP}}\)-absolutely continuous part of
any \(\mQ\in\pmea{\outA}\)   by
\(\qmn{ac}\). Then\vspace{-.1cm}
\begin{align}
\notag
\hspace{0.7cm}&\hspace{-0.7cm}
\CRD{\rno}{\Wm}{\mQ}{\mP}
\!-\!
\CRD{\rno}{\Wm}{\qmn{\rno,\mP}}{\mP}
\\
\notag
&\stackrel{(a)}{\leq}\CRD{\rno}{\Wm}{\qmn{ac}}{\mP}
\!-\!
\CRD{\rno}{\Wm}{\qmn{\rno,\mP}}{\mP}
\\
\notag
&=\tfrac{1}{\rno-1}\EXS{\mP\!}{\ln \int  
	\left(\der{\qmn{ac}}{\qmn{\rno,\mP}}\right)^{1-\rno}\Wma{\rno}{\qmn{\rno,\mP}}(\dif{\dout}|\inp)}
\\
\notag
&\stackrel{(b)}{\leq}
\begin{cases}
\tfrac{1}{\rno-1}\EXS{\mP\!}{ \displaystyle{\int} \ln  
	\left(\der{\qmn{ac}}{\qmn{\rno,\mP}}\right)^{1-\rno}\Wma{\rno}{\qmn{\rno,\mP}}(\dif{\dout}|\inp)}
&\text{if~}\rno\!<\!1
\\
\tfrac{1}{\rno-1}\ln\EXS{\mP\!}{\displaystyle{\int}
	\left(\der{\qmn{ac}}{\qmn{\rno,\mP}}\right)^{1-\rno}\Wma{\rno}{\qmn{\rno,\mP}}(\dif{\dout}|\inp)}
&\text{if~}\rno\!>\!1
\end{cases} 
\\
\notag
&\stackrel{(c)}{=}
\RD{1\vee \rno}{\qmn{\rno,\mP}}{\qmn{ac}}
\\
\notag
&\stackrel{(d)}{=}
\RD{1\vee \rno}{\qmn{\rno,\mP}}{\mQ},
\end{align}
where \((a)\) follows from \(\qmn{ac}\!\leq\!\mQ\) and \cite[Lemma 1]{nakiboglu19C},
\((b)\) follows from Jensen's inequality and
the concavity of natural logarithm function,
\((c)\)  follows from \eqref{eq:def:Aoperator},
\eqref{eq:lem:fixedpoint}, and 
Fubini's theorem \cite[Thm. 3.4.4]{bogachev},
and \((d)\) follows from the definition of \renyi divergence.

\bibliographystyle{IEEEtran}
\balance
\newcommand{\noopsort}[1]{} \newcommand{\printfirst}[2]{#1}
  \newcommand{\singleletter}[1]{#1} \newcommand{\switchargs}[2]{#2#1}

\end{document}